\documentclass[preprint2]{aastex}

\shorttitle{Evidence for long-term Gamma-ray and X-ray variability from the unidentified TeV source HESS J0632+057}
\shortauthors{Maier et al.}

\begin{document}
\title{Evidence for long-term Gamma-ray and X-ray variability from the unidentified TeV source HESS J0632+057}

\author{
V. A. Acciari\altaffilmark{1},
E. Aliu\altaffilmark{2},
T. Arlen\altaffilmark{3},
M. Beilicke\altaffilmark{4},
W. Benbow\altaffilmark{1},
D. Boltuch\altaffilmark{2},
S. M. Bradbury\altaffilmark{5},
J. H. Buckley\altaffilmark{4},
V. Bugaev\altaffilmark{4},
K. Byrum\altaffilmark{6},
A. Cannon\altaffilmark{7},
A. Cesarini\altaffilmark{8},
A. Cesarini\altaffilmark{8},
Y. C. Chow\altaffilmark{3},
L. Ciupik\altaffilmark{9},
P. Cogan\altaffilmark{10},
R. Dickherber\altaffilmark{4},
C. Duke\altaffilmark{11},
T. Ergin\altaffilmark{12},
A. Falcone\altaffilmark{13},
S. J. Fegan\altaffilmark{3},
J. P. Finley\altaffilmark{14},
G. Finnegan\altaffilmark{15},
P. Fortin\altaffilmark{16},
L. Fortson\altaffilmark{9},
A. Furniss\altaffilmark{17},
K.Gibbs\altaffilmark{1},
G. H. Gillanders\altaffilmark{8},
J. Grube\altaffilmark{7},
R. Guenette\altaffilmark{10},
G. Gyuk\altaffilmark{9},
D. Hanna\altaffilmark{10},
J. Holder\altaffilmark{2},
D. Horan\altaffilmark{18},
C. M. Hui\altaffilmark{15},
T. B. Humensky\altaffilmark{19},
A. Imran\altaffilmark{20},
P. Kaaret\altaffilmark{21},
N. Karlsson\altaffilmark{9},
M. Kertzman\altaffilmark{22},
D. Kieda\altaffilmark{15},
J. Kildea\altaffilmark{1},
A. Konopelko\altaffilmark{23},
H. Krawczynski\altaffilmark{4},
F. Krennrich\altaffilmark{20},
M. J. Lang\altaffilmark{8},
S. LeBohec\altaffilmark{15},
G. Maier\altaffilmark{10,*},
A. McCann\altaffilmark{10},
M. McCutcheon\altaffilmark{10},
J. Millis\altaffilmark{24},
P. Moriarty\altaffilmark{25},
R. Mukherjee\altaffilmark{16},
R. A. Ong\altaffilmark{3},
A. N. Otte\altaffilmark{17},
D. Pandel\altaffilmark{21},
J. S. Perkins\altaffilmark{1},
D. Petry\altaffilmark{26},
M. Pohl\altaffilmark{20},
J. Quinn\altaffilmark{7},
K. Ragan\altaffilmark{10},
L. C. Reyes\altaffilmark{27},
P. T. Reynolds\altaffilmark{28},
H. J. Rose\altaffilmark{5},
M. Schroedter\altaffilmark{20},
G. H. Sembroski\altaffilmark{14},
A. W. Smith\altaffilmark{6},
D. Steele\altaffilmark{9},
M. Theiling\altaffilmark{1},
J. A. Toner\altaffilmark{8},
A. Varlotta\altaffilmark{14},
V. V. Vassiliev\altaffilmark{3},
S. Vincent\altaffilmark{15},
R. G. Wagner\altaffilmark{6},
S. P. Wakely\altaffilmark{19},
J. E. Ward\altaffilmark{7},
T. C. Weekes\altaffilmark{1},
A. Weinstein\altaffilmark{3},
T. Weisgarber\altaffilmark{19},
D. A. Williams\altaffilmark{17},
S. Wissel\altaffilmark{19},
M. Wood\altaffilmark{3}
}

\altaffiltext{*}{Corresponding author; gernot.maier@mcgill.ca}
\altaffiltext{1}{Fred Lawrence Whipple Observatory, Harvard-Smithsonian Center for Astrophysics, Amado, AZ 85645, USA}
\altaffiltext{2}{Department of Physics and Astronomy and the Bartol Research Institute, University of Delaware, Newark, DE 19716, USA}
\altaffiltext{3}{Department of Physics and Astronomy, University of California, Los Angeles, CA 90095, USA}
\altaffiltext{4}{Department of Physics, Washington University, St. Louis, MO 63130, USA}
\altaffiltext{5}{School of Physics and Astronomy, University of Leeds, Leeds, LS2 9JT, UK}
\altaffiltext{6}{Argonne National Laboratory, 9700 S. Cass Avenue, Argonne, IL 60439, USA}
\altaffiltext{7}{School of Physics, University College Dublin, Belfield, Dublin 4, Ireland}
\altaffiltext{8}{School of Physics, National University of Ireland, Galway, Ireland}
\altaffiltext{9}{Astronomy Department, Adler Planetarium and Astronomy Museum, Chicago, IL 60605, USA}
\altaffiltext{10}{Physics Department, McGill University, Montreal, QC H3A 2T8, Canada}
\altaffiltext{11}{Department of Physics, Grinnell College, Grinnell, IA 50112-1690, USA}
\altaffiltext{12}{Harvard-Smithsonian Center for Astrophysics, 60 Garden Street, Cambridge, MA 02138, USA}
\altaffiltext{13}{Department of Astronomy and Astrophysics, 525 Davey Lab, Pennsylvania State University, University Park, PA 16802, USA}
\altaffiltext{14}{Department of Physics, Purdue University, West Lafayette, IN 47907, USA }
\altaffiltext{15}{Physics Department, University of Utah, Salt Lake City, UT 84112, USA}
\altaffiltext{16}{Department of Physics and Astronomy, Barnard College, Columbia University, NY 10027, USA}
\altaffiltext{17}{Santa Cruz Institute for Particle Physics and Department of Physics, University of California, Santa Cruz, CA 95064, USA}
\altaffiltext{18}{Laboratoire Leprince-Ringuet, Ecole Polytechnique, CNRS/IN2P3, F-91128 Palaiseau, France}
\altaffiltext{19}{Enrico Fermi Institute, University of Chicago, Chicago, IL 60637, USA}
\altaffiltext{20}{Department of Physics and Astronomy, Iowa State University, Ames, IA 50011, USA}
\altaffiltext{21}{Department of Physics and Astronomy, University of Iowa, Van Allen Hall, Iowa City, IA 52242, USA}
\altaffiltext{22}{Department of Physics and Astronomy, DePauw University, Greencastle, IN 46135-0037, USA}
\altaffiltext{23}{Department of Physics, Pittsburg State University, 1701 South Broadway, Pittsburg, KS 66762, USA}
\altaffiltext{24}{Department of Physics, Anderson University, 1100 East 5th Street, Anderson, IN 46012}
\altaffiltext{25}{Department of Life and Physical Sciences, Galway-Mayo Institute of Technology, Dublin Road, Galway, Ireland}
\altaffiltext{26}{European Southern Observatory, Karl-Schwarzschild-Strasse 2, 85748 Garching, Germany}
\altaffiltext{27}{Kavli Institute for Cosmological Physics, University of Chicago, Chicago, IL 60637, USA}
\altaffiltext{28}{Department of Applied Physics and Instrumentation, Cork Institute of Technology, Bishopstown, Cork, Ireland}

\begin{abstract}
HESS J0632+057 is one of only two unidentified very-high-energy gamma-ray sources which appear to be point-like within experimental resolution. 
It is possibly associated with the massive Be star MWC 148 and has been suggested to resemble known
TeV binary systems like LS I +61 303 or LS 5039. 
HESS J0632+057 was observed by VERITAS 
for  31 hours in 2006, 2008 and 2009.
During these observations, no significant signal in gamma rays with energies above 1 TeV was detected from the direction of HESS J0632+057.
A flux
upper limit corresponding to 1.1\% of the flux of the Crab Nebula has been derived from the VERITAS data.
The non-detection by VERITAS excludes  with a probability of 99.993\% that HESS J0632+057 is a steady gamma-ray emitter.  
Contemporaneous X-ray observations with \emph{Swift} XRT reveal a factor of $1.8 \pm 0.4$ higher flux
 in the 1-10 keV range than earlier X-ray observations of HESS J0632+057.
The variability in the gamma-ray and X-ray fluxes supports interpretation of the object as a gamma-ray emitting binary.
\end{abstract}


\section{Introduction}

HESS J0632+057 is one of about 20 very-high-energy (VHE) gamma-ray sources with no known counterparts at other
wavelengths (for a recent review see e.g.~\cite{Weekes-2008}).
Gamma-ray emission was discovered by the High Energy Stereoscopic System (H.E.S.S.) 
during observations of the Monoceros Loop Supernova Remnant in 2004 and 2005 \citep{Aharonian-2007}.
It appears to be point-like within experimental resolution; the limit on the size of the emission region was given as 2' (95\%  confidence level).
The reported flux of gamma rays with energies above 1 TeV from HESS J0632+057 corresponds to about 3\% of the flux of the Crab
Nebula, with a differential photon spectrum  consistent with a power-law function with 
index of $2.53\pm0.26_{\mathrm{stat}}\pm0.20_{\mathrm{sys}}$.
Possible associations considered by \cite{Aharonian-2007} are the Monoceros Loop Supernova remnant, the weak X-ray source 1RXS J063258.3+054857,
the B0pe-star MWC 148 (HD 259440) and the unidentified GeV gamma-ray source 3EG J0634+0521 \citep{Hartman-1999}.
Follow-up X-ray observations with \emph{XMM-Newton} by \cite{Hinton-2009} revealed a variable X-ray source
(XMMU J063259.3+054801) with a position compatible with HESS J0632+057 and MWC 148.
It should be noted that 3EG J0634+0521 is absent in the EGR catalogue of EGRET gamma-ray sources, a reanalysis of the
EGRET data with new Galactic interstellar emission models and interstellar radiation field data \citep{Casandjian-2008}.
3EG J0634+0521 is, as expected from the reported flux, not in the Fermi bright gamma-ray source list \citep{Abdo-2009}.

Point-like gamma-ray sources stand out among the many galactic VHE objects with spatially extended gamma-ray emission.
The latter are usually associated with either pulsar wind nebulae or supernova remnants.
High-mass X-ray binaries constitute the only known class of galactic objects with variable point-like VHE emission; this class 
 currently contains three members only: PSR B1959-63/SS 2883 \citep{Aharonian-2005b}, LS 5039 \citep{Aharonian-2005c} and
LS I +61 303 \citep{Albert-2006, Acciari-2008}.
Additionally, marginal evidence at a level of 3.2$\sigma$ from the black-hole binary Cyg X-1 has been reported by \cite{Albert-2007}.
TeV binaries show variable emission of gamma rays, likely connected to changes in physical parameters 
associated with the orbital movement.
VHE gamma-ray production in these objects is explained by the acceleration of charged particles in
accretion-powered relativistic jets \citep{Taylor-1984,Mirabel-1994}
or in shocks created by
the collision of the expanding pulsar wind with the wind from the stellar companion \citep{Maraschi-1981}.
Subsequent inverse-Compton scattering on low-energy stellar photons produces gamma rays at GeV and TeV energies.
While there has been no compact companion discovered for MWC 148, the point-like nature of the VHE
emission combined with the variable X-ray emission can easily be explained by a production 
scenario similar to those in TeV binaries.
A second possible scenario is that MWC 148 is a representative of a new type of VHE emitter as proposed by
\cite{Babel-1997} and \cite{Townsend-2007}.
In their picture strong magnetic fields around the massive star lead to magnetically channeled wind shocks in which
second-order Fermi acceleration might occur.
However it is not clear if the circumstellar environment of MWC 148 is strongly magnetized,
or if this acceleration mechanism is able to produce particles of sufficiently high energy to produce a measurable TeV flux.
An association of HESS J0632+057 with the Monoceros loop SNR is unlikely given the point-like nature of the
gamma-ray emission and the non-correlation of possible target material with the position of the VHE source 
\citep{Aharonian-2007}.


\section{Observations}

VERITAS is an array of four imaging atmospheric-Cherenkov telescopes
located at the Fred Lawrence Whipple Observatory
in southern Arizona.
It combines a large effective
area (up to $10^5$ m$^2$) over a wide energy range (100 GeV
to 30 TeV) with good energy (15-20\%) and angular
($\approx 0.1^{\mathrm{o}}$) resolution.
The field of view of the VERITAS telescopes is 3.5$^{\mathrm{o}}$.
The high sensitivity of VERITAS enables
the detection of sources with a flux of 1\% of the Crab
Nebula in less than 50 hours of observations.
For more details on the VERITAS
instrument, see e.g.~\cite{Acciari-2008}.

VERITAS observed the sky around HESS J0632+057 during three periods 
 in
December 2006, December 2008 and January 2009; see Table \ref{table:observations} 
for details.
For each period, data equivalent to 10 hours of observations passed quality selection criteria,
which remove data taken during bad weather or with
hardware-related problems.
Data were taken on moonless nights 
in wobble mode, wherein
the source was positioned 0.5 degrees from the camera center with the offsets in
different positions for different runs.
The first data set (Set I) consists of observations taken
during the construction phase of VERITAS
with only 3 telescopes.
These observations were pointed towards the centre of the Monoceros region (at an angular distance of $\sim0.5^{\mathrm{o}}$ from
HESS J0632+057),
while observations in
the second and third set were targeted around the reported
position of HESS J0632+057.

The data analysis steps consist of image calibration and cleaning,
second-moment parameterization of these images \citep{Hillas-1985},
stereoscopic reconstruction of the event impact position and direction,
gamma-hadron separation, spectral energy reconstruction
(see e.g.~\cite{Krawczynski-2006})
as well as the generation of photon sky maps.
The majority of the far more numerous background events are rejected by comparing the shape of the event images in each 
telescope with the expected shapes of gamma-ray showers modeled by Monte Carlo simulations.
These \textit{mean-reduced-scaled width} and \textit{mean-reduced-scaled length}
cuts (see definition in \cite{Krawczynski-2006}), and an additional cut on the arrival direction of the incoming gamma ray ($\Theta^2$)
reject more than 99.9\% of the cosmic-ray background while keeping 45\% of the gamma rays.
The cuts applied here are:
integrated charge per image $>$ 1200 digital counts ($\approx$225 photoelectrons), 
$-1.2 < $ mean-reduced-scaled width/length $< 0.5$, and \mbox{$\Theta^2 < 0.015$ deg$^2$}
(\mbox{$\Theta^2 < 0.025$ deg$^2$} for the 3-telescope data set).
The background in the source region is estimated from the same field of view
using the ``reflected-region'' model with 8-10 background regions \citep{Aharonian-2001}
 and the ``ring-background'' model  with a ring size of 0.5$^{\mathrm{o}}$ (mean radius) and a ring width
 of 0.175$^{\mathrm{o}}$ \citep{Aharonian-2005a}. 
This analysis has been chosen to provide best sensitivity at high energies, in order to compare more directly with the
results reported in \cite{Aharonian-2007}.
The resulting energy threshold is 720 GeV.


Observations of XMMU\,J063259.3+054801 (HESS J0632+057) were taken  with the \emph{Swift}
satellite over four consecutive nights, from January 26 to January 29, 2009, and were
contemporaneous with VERITAS observations (Set III). 
The analysis is restricted to
X-ray data from the XRT instrument \citep{Burrows-2005} since the
source is not detected with BAT and the bright star 
MWC 148
causes a
very high level of photon-coincidence losses in the UVOT instrument 
\citep{Poole-2008}.\footnote{It should be noted that MWC 148 with a B magnitude of less than 9 is too faint to introduce
systematic effects into the analysis of the VERITAS data.}
The presented \emph{Swift} XRT observations were conducted
in photon-counting (PC) mode with no signatures of pile-up. Data
reduction is performed with the HEAsoft 6.5 package. Events with grades
0--12 in the energy range of 0.3--10 keV are calibrated and cleaned
using the \emph{xrtpipeline} tool by applying the standard filtering
criteria and latest \emph{Swift} calibration files. Source counts are
extracted from a circular region with a 30-pixel radius (47.2 arcsec),
and background events are extracted from a 40-pixel radius circle in a
source-free region. Ancillary response files are generated with the
\emph{xrtmkarf} tool applying corrections for the PSF losses and CCD
defects. The latest response matrix from the XRT calibration files are
used in the analysis. To ensure valid $\chi^{2}$ minimization statistics
during spectral fitting, the extracted XRT energy spectra are
rebinned to contain a minimum of 25 counts in each bin.

\section{Results}

\begin{figure}
\includegraphics[width=0.95\linewidth]{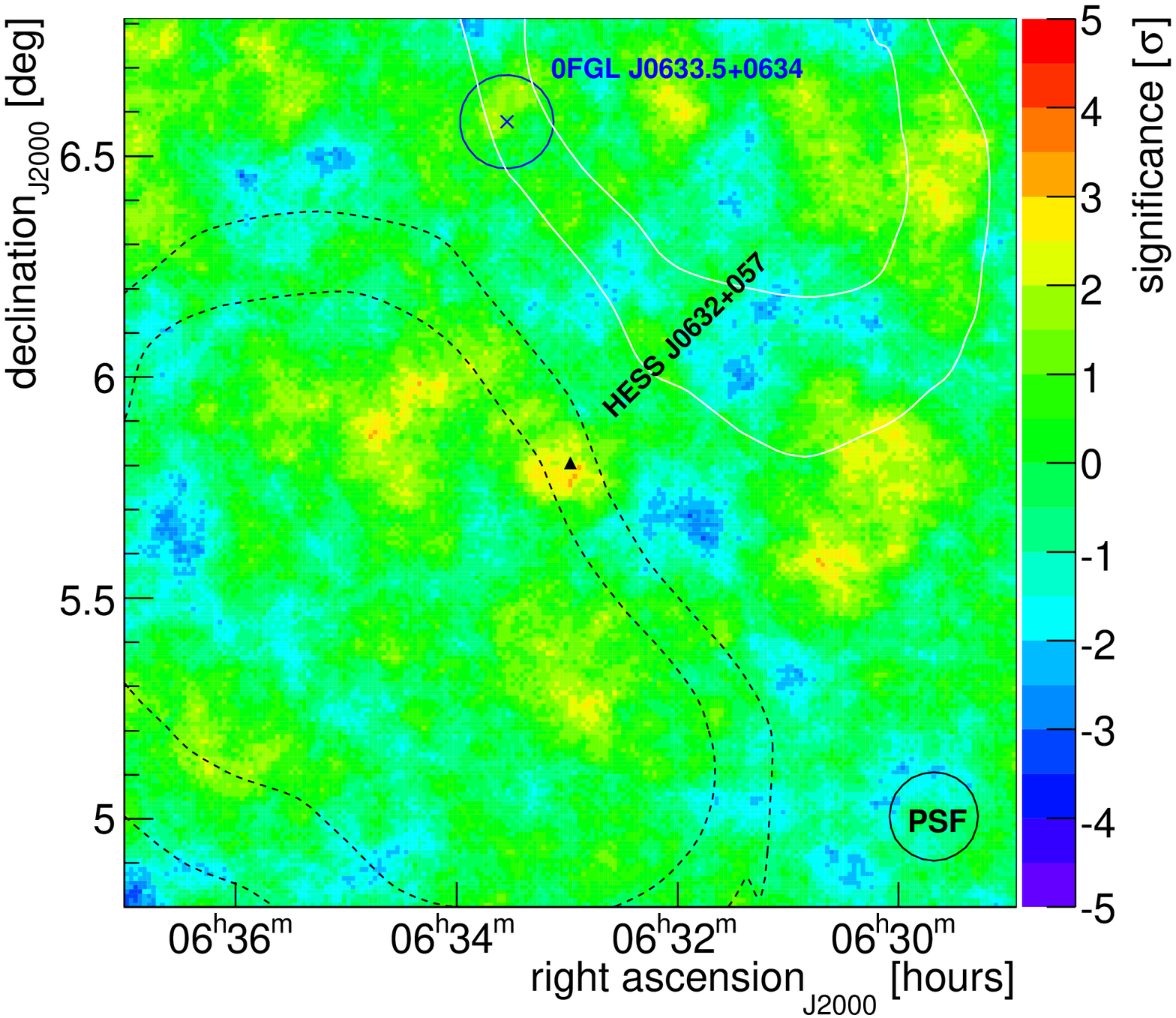}
\caption{\label{fig-skyPlot}
VERITAS significance map of the region around HESS J0632+057 for the whole data set
and an energy threshold of 720 GeV.
The background is estimated using
the ring background method.
The location of HESS J0632+057 is indicated by a black triangle.
Also shown are the 95\% and 99\% confidence regions of the EGRET 
sources 3EGJ0634+0521 (white lines) and 3EGJ0631+0642 (GeV J0633+0645) (black lines).
The Fermi source 0FGLJ0633.5+0634 is indicated with a `x' sign; the blue
circle denotes the 95\% confidence region \citep{Abdo-2009}.
The circle at the bottom right indicates the angular resolution of the VERITAS observations.
}
\end{figure}

\begin{figure}
\plotone{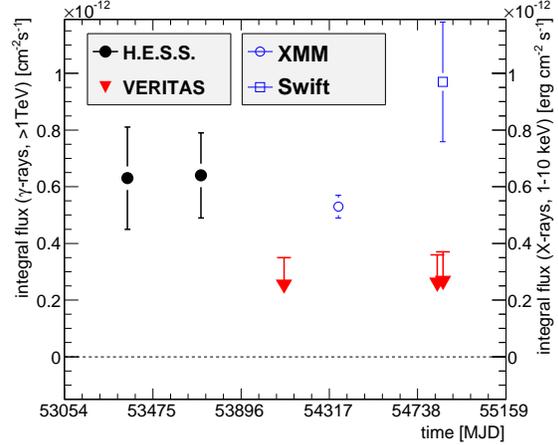}
\caption{\label{fig-lightcurve} 
The light curve above 1 TeV from HESS J0632+057 is shown assuming a spectral
shape of $dN/dE \propto E^{-\Gamma}$ with $\Gamma=2.5$. 
The downwards pointing arrows show the 99\% confidence limits derived here from the VERITAS data.
The H.E.S.S. fluxes are taken from \cite{Aharonian-2007}.
The X-ray fluxes measured by \emph{XMM-Newton} and \emph{Swift} are indicated by open symbols.
}
\end{figure}

Results for each of the three VERITAS data sets (see Table \ref{table:observations}), as well as for the total
observation are listed in Table 2.  
Figure \ref{fig-skyPlot}  shows a sky map of the significance at energies above 720 GeV observed in the region
around HESS J0632+057.
The distribution of significances in the sky map is consistent with the expected distribution from
a field with
no gamma-ray source present. The significance at the position of HESS J0632+057 is $2.1\sigma$ ($1\sigma$ for
an energy threshold of 1 TeV, see Table \ref{table:results}).
There is therefore no significant evidence for gamma-ray like events from HESS J0632+057  during the 31 hours of observations
with VERITAS.
The flux upper limit (E$>$1 TeV) for the complete data set at the 99\% confidence level \citep{Helene-1983}
assuming a power-law like source spectrum with a spectral index of $\Gamma=2.5$ is
 F($>$1 TeV) $< 2.6\times 10^{-13}$ cm$^{-2}$s$^{-1}$ (about 1.1\% of the flux of the
Crab Nebula; see Table \ref{table:results}).
This flux limit is $\sim2.4$ times lower than the flux reported by H.E.S.S in \cite{Aharonian-2007};
see Figure \ref{fig-lightcurve} for a light curve.
The probability for a non-variable flux of high-energy gamma rays from HESS J0632+057 is 
derived from the VERITAS data
 and the average of the reported fluxes from H.E.S.S. using a $\chi^2$-test. 
The test gives a $\chi^2$ of 15.8 with 1 degree of freedom, corresponding to a probability of 0.007\% (about 4$\sigma$).


The non-detection of HESS J0632+057 by VERITAS initiated additional and very thorough
quality checks of the data.
Optical pointing monitors which are installed on each telescope show that the systematic error on the pointing
was less than 90'' during the data-taking period for Set II and III\footnote{There were no pointing monitors installed
in December 2006 (Set I).}.
The sensitivity of VERITAS in the elevation range of 50-60$^{\mathrm{o}}$ was confirmed by observations of the Crab Nebula and
several other weak gamma-ray sources in early 2009 (e.g. \cite{Ong-2009}).

The \emph{Swift} XRT observations provide further evidence for X-ray flux variability in the
object. 
The total photon spectrum
from the four \emph{Swift} XRT observations in January 2009 is well-described 
by an absorbed power law. Due to the low counting statistics,
a fixed column density of $\rm{N}_{\rm{H}} = 3.1 \times 10^{21}$ cm
$^{-2}$ is applied in the spectral analysis, as measured for the
\emph{XMM-Newton} EPIC spectrum \citep{Hinton-2009}. The best-fit
absorbed power law model yields a $\chi^{2}/\rm{dof} = 3.5/7$, a photon
index $\Gamma = 1.9 \pm 0.3$, and a 1 keV normalization of $(2.4 \pm
0.6) \times 10^{-4}$ cm$^{-2}$ s$^{-1}$ keV$^{-1}$. The deabsorbed 1-10
keV flux is $(9.7 \pm 2.1) \times 10^{-13}$ erg cm$^{-2}$ s$^{-1}$,
corresponding to a factor of $1.8 \pm 0.4$ higher flux than the
\emph{XMM-Newton} EPIC data. Among the four \emph{Swift} XRT
observations no significant flux variability is measured. Figure 3 shows
the deabsorbed X-ray spectra from the \emph{Swift} and
\emph{XMM-Newton} observations. 
Over the limited energy range of 0.8--4
keV,  the \emph{Swift} photon spectra is softer 
($\Gamma = 1.9 \pm 0.3$) than the  \emph{XMM-Newton} EPIC spectrum ($\Gamma = 1.26 \pm 0.04$).
The relatively hard ($\Gamma
\leq 2$) X-ray spectrum is similar to the spectral slopes measured from
the TeV binaries LS\,I\,+61\,303 \citep{Leahy-1997},  LS\,5039
\citep{Hoffmann-2009}, and PSR B1259-63/SS 2883 \citep{Chernyakova-2006}.
No definite conclusion can be drawn from the X-ray spectral variability of  HESS J0632+057.
It appears, for example from observations of PSR B1259-63/SS 2883 \citep{Chernyakova-2006} that 
the photon index of the X-ray spectra in TeV binaries is not 
a strictly monotonic function of flux. 
 Detailed results from the full \emph{Swift}
observing campaign of XMMU\,J063259.3+054801 between January 2009 and
April 2009 are presented in \cite{Falcone-2009}.

The non-detection of HESS J0632+057 by VERITAS provides evidence for variability in the flux of gamma-rays with
energies above 1 TeV. 
HESS J0632+057 is unlikely to have a blazar counterpart. 
Blazars are generally hard X-ray point sources and bright radio sources. 
Neither the XMM observation nor the Swift images show the presence of any bright
 X-ray source other than the Be star within the HESS error circle. 
 A search in the NVSS catalog also finds no radio source in the area.
The VHE emission and variability can be easily explained if MWC 148 would be part of a binary system and high-energy photons
are produced in a similar way as in LS I +61 303 or LS 5039.
The available data do not allow any conclusion on a possible periodicity of the gamma-ray signal. 
A detection of the compact companion or of the orbital motion of the Be star is required to confirm or refute the binary nature of this system.
Particle acceleration and VHE emission from massive stars with strong magnetic fields has also been suggested.
A confirmation that MWC 148//HESS J0632+057 is surrounded by sufficiently strong magnetic fields, along with further theoretical work
to explain the variability in the gamma-ray emission, would be needed to establish this potentially new class of
galactic gamma-ray sources.
Future multiwavelength observations combined with results from ground-based and space-based gamma-ray observatories
will provide a deeper understanding of the true nature of HESS J0632+057.

\acknowledgments

This research is supported by grants from the US Department
of Energy, the US National Science Foundation, and the Smithsonian
Institution, by NSERC in Canada, by Science Foundation
Ireland, and by STFC in the UK.
We acknowledge the
excellent work of the technical support staff at the FLWO and
the collaborating institutions in the construction and operation
of the instrument.
We acknowledge the efforts of the Swift team
for providing the UVOT/XRT observations.

{\it Facilities:} \facility{VERITAS}, \facility{Swift}

\clearpage

\begin{figure}
\includegraphics[width=0.95\linewidth]{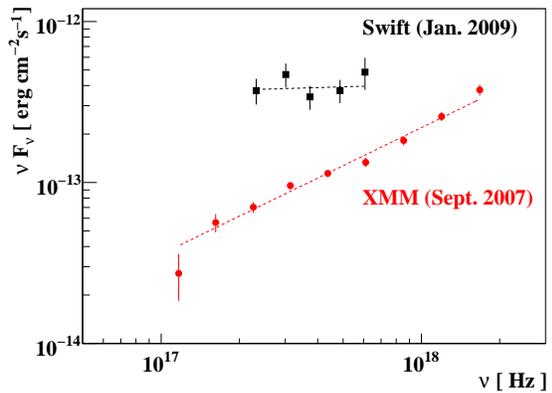}
\caption{\label{fig:SwiftSpectra}
X-ray spectrum of HESS J0632+057 from \emph{XMM-Newton} EPIC data in September 2007 \citep{Hinton-2009}
and \emph{Swift} XRT observations in January 2009.
}
\end{figure}

\clearpage

\begin{deluxetable}{llcccc}
\centering
\tablecolumns{6}
\tablewidth{0pt}
\tablecaption{
Details of the VERITAS and H.E.S.S. \citep{Aharonian-2007} observations of \mbox{HESS J0632+057}. 
\label{table:observations}}
\tablehead{
\colhead{Name} &
\colhead{Date range} &
\colhead{N$_{\mathrm{Tel}}$\tablenotemark{a}} &
\colhead{Elevation} &
\colhead{Angular distance} &
\colhead{Observation} \\
\colhead{} &
\colhead{} &
\colhead{} &
\colhead{range} &
\colhead{between source and} &
\colhead{time} \\
\colhead{} &
\colhead{} &
\colhead{} &
\colhead{} &
\colhead{pointing direction} &
\colhead{ [min] } 
}
\startdata
Set I  & Dec 16 2006 - Jan 25 2007 & 3 & 55-65$^{\mathrm{o}}$ & 0.15-0.8$^{\mathrm{o}}$ & 580  \\
Set II  & Dec 30 2008 - Jan 03 2009 & 4 & 59-65$^{\mathrm{o}}$ & 0.5$^{\mathrm{o}}$ & 560  \\
Set III  & Jan 26 2009 - Jan 30 2009 & 3/4 & 59-65$^{\mathrm{o}}$ & 0.5$^{\mathrm{o}}$ & 722 \\
\hline
H.E.S.S. P1 & Dec 2004 & & & & 282 \\
H.E.S.S. P2 & Nov 2005 - Dec 2005 & & & & 372 \\
\enddata 
\tablenotetext{a}{Number of available telescopes}
\end{deluxetable}

\begin{deluxetable}{lcccccc}
\centering
\tablecolumns{6}
\tablewidth{0pt}
\tablecaption{
Analysis results for \mbox{HESS J0632+057} for $E>1$ TeV.
Upper limits $\Phi_{\gamma, \mathrm{UL}} (E>1$ TeV$)$
are given at 99\% confidence level 
(after \cite{Helene-1983}).
The integral fluxes and 1 $\sigma$ errors above 1 TeV reported by H.E.S.S. are listed for comparison \citep{Aharonian-2007}.
\label{table:results}}
\tablehead{
\colhead{} &
\colhead{{\it on}} &
\colhead{{\it off}} &
\colhead{$\alpha$} &
\colhead{excess} &
\colhead{significance} &
\colhead{Flux or upper flux limit}\\
\colhead{} &
\colhead{events} &
\colhead{events} &
\colhead{} &
\colhead{events} &
\colhead{ [$\sigma$] } &
\colhead{ [$10^{-13}$ cm$^{-2}$s$^{-1}$] } 
}
\startdata
Set I & 84 & 594 & 0.16 & -8.8 & -0.9 & $<4.2$ \\
Set II & 131 & 713 & 0.16 & 15.7 & 1.3 & $<4.2$ \\
Set III & 120 & 669 & 0.16 & 11.4 & 1.0 & $<$ 3.6 \\
\hline
Total (I-III) & 335 & 1976 & 0.16 & 19.4 & 1.0 & $<$ 2.6 \\
\hline
H.E.S.S. P1 & & & & & & $6.3\pm1.8$ \\ 
H.E.S.S. P2 & & & & & & $6.4\pm1.5$\\
\enddata 
\end{deluxetable}


\begin{thebibliography}{}
\bibitem[Abdo et al(2009)]{Abdo-2009} Abdo, A.A. et al (Fermi collaboration) 2009, submitted to \apjs, astro-ph/0902.1340
\bibitem[Acciari et al(2008)]{Acciari-2008} Acciari, V.A. et al (VERITAS collaboration) 2008, \apj, 679, 1427
\bibitem[Aharonian et al(2001)]{Aharonian-2001}  Aharonian, F. et al (HEGRA collaboration) 2001, \aap, 370, 112    
\bibitem[Aharonian et al(2005a)]{Aharonian-2005a} Aharonian, F. et al (H.E.S.S. collaboration) 2005, \aap, 430, 865
\bibitem[Aharonian et al(2005b)]{Aharonian-2005b} Aharonian, F. et al (H.E.S.S. collaboration) 2005, \aap, 442, 1
\bibitem[Aharonian et al(2005c)]{Aharonian-2005c} Aharonian, F. et al (H.E.S.S. collaboration) 2005, Science, 309, 746
\bibitem[Aharonian et al(2007)]{Aharonian-2007} Aharonian, F. et al (H.E.S.S. collaboration) 2007, \aap, 469, L1
\bibitem[Albert et al(2006)]{Albert-2006} Albert, J. et al (MAGIC collaboration) 2006, Science, 312, 1771
\bibitem[Albert et al(2007)]{Albert-2007} Albert, J. et al (MAGIC collaboration) 2007, \apj 665, L51
\bibitem[Babel \& Montmerle(1997)]{Babel-1997} Babel, J. \& Montmerle, T. 1997, \aap, 323, 121
\bibitem[Burrows et al(2005)]{Burrows-2005} Burrows D. et al 2005, SSRv, 120, 165
\bibitem[Casandjian \& Grenier(2008)]{Casandjian-2008} Casandjian, J.-M. \& Grenier, I. A. 2008 \aap, 489, 849
\bibitem[Chernyakova et al(2006)]{Chernyakova-2006} Chernyakova, M. et al 2006, \mnras 367, 1021
\bibitem[Falcone et al(2009)]{Falcone-2009} Falcone, A. et al 2009, in preparation
\bibitem[Helene(1983)]{Helene-1983} Helene, O. 1983, Nuclear Instruments \& Methods 212, 319
\bibitem[Hartman et al(1999)]{Hartman-1999} Hartman, R.C. et al 1993, \apjs, 123, 79 
\bibitem[Hillas(1985)]{Hillas-1985} Hillas M. 1985, Proc. of the 19th ICRC (La Jolla, USA), 3, 445
\bibitem[Hinton et al(2009)]{Hinton-2009} Hinton, J. et al 2009, \apjl, 690, L101
\bibitem[Hoffmann et al(2009)]{Hoffmann-2009} Hoffmann, A.D. et al 2009, \aap, 494, L37
\bibitem[Krawczynski et al.(2006)]{Krawczynski-2006} Krawczynski H. et al. 2006, Astroparticle Physics, 25, 380
\bibitem[Leahy et al(1997)]{Leahy-1997} Leahy, D.A. et al 1997, \apj, 475, 823
\bibitem[Maraschi \& Treves(1981)]{Maraschi-1981} Maraschi, L. \& Treves, A. 1981, \mnras, 194, 1
\bibitem[Mirabel \& Rodriguez(1994)]{Mirabel-1994} Mirabel, I. \& Rodriguez, L.F. 1994, \nat, 371, 46
\bibitem[Ong et al(2009)]{Ong-2009} Ong, R. et al (VERITAS collaboration) 2009, Astronomer's Telegram \#1941
\bibitem[Poole et al(2008)]{Poole-2008} Poole T.S. et al 2008, \mnras 383, 627 
\bibitem[Taylor \& Gregory(1984)]{Taylor-1984} Taylor, A.R. \& Gregory P.C. 1984, \apj, 283, 273
\bibitem[Townsend et al(2007)]{Townsend-2007} Townsend, R.H.D., Owocki, S.P. \& ud-Doula, A. 2007, \mnras, 382, 139
\bibitem[Weekes(2008)]{Weekes-2008} Weekes, T. 2008, astro-ph/0811.1197
\end{thebibliography}
\end{document}